\documentclass[10pt,conference]{IEEEtran}
\usepackage{cite}
\usepackage{amsmath,amssymb,amsfonts}
\usepackage{algorithm}
\usepackage{algorithmic}
\usepackage{graphicx}
\usepackage{textcomp}
\usepackage{xcolor}
\usepackage[hyphens]{url}
\usepackage{hyperref}
\usepackage{booktabs}
\usepackage{multirow}
\usepackage{caption}
\hypersetup{hidelinks,
  pdfauthor={Jingyuan Li, Maolin Wang, Hayden So},
  pdftitle={A Per-Step Load-Balancing Engine for Expert-Parallel MoE Serving}}
\graphicspath{{figures/}}

\title{A Per-Step Load-Balancing Engine\\
for Expert-Parallel MoE Serving}

\author{%
  \IEEEauthorblockN{Jingyuan Li\textsuperscript{1}, Maolin Wang\textsuperscript{2}, and Hayden So\textsuperscript{1}}
  \IEEEauthorblockA{\textsuperscript{1}The University of Hong Kong\\
                   \textsuperscript{2}The Hong Kong University of Science and Technology}%
}

\newcommand{\ueff}{\ensuremath{U_{\mathrm{eff}}}}
\newcommand{\tmoe}{\ensuremath{T_{\mathrm{moe}}}}
\newcommand{\tsolve}{\ensuremath{T_{\mathrm{solve}}}}

\begin{document}
\maketitle
\thispagestyle{plain}
\pagestyle{plain}

\begin{abstract}
In expert-parallel Mixture-of-Experts serving, a layer finishes only when
the most loaded GPU of its group does. Token routing shifts at every
decode step and concentrates on a few experts, so most of the group's
reserved GPU time waits on stragglers. Recovering that time means
re-deciding, at every layer of every decode step, how each replicated
expert's tokens divide between its copies. The exact linear-program
solvers that make this decision optimally take longer than many of the
layers they serve, and their working set outgrows GPU shared memory at
deployment scale. We introduce Relief, a bidirectional pairwise
relaxation that makes the linear program unnecessary, coming within a
fraction of a point of the per-step optimum in a few passes of integer
halving over the token counts. We build and measure three deployments, CUDA kernels on the
serving GPU, the same verified engine on an FPGA over PCIe, and a
fixed-function block beside the accelerator. On decode routing from
three open MoE families, Relief holds the balance near the per-step
optimum at every group size and layer time measured. Scored as effective
utilization relative to that optimum, the kernel alone matches or
exceeds every deployable scheme at every group size, from the
hundred-microsecond layer on. The on-die engine widens that lead at every point, recovering up
to 37.3 percentage points over the deployed static placement, while adding 0.02\% of a current accelerator's transistors.
\end{abstract}

\section{Introduction}
\label{sec:intro}

Mixture-of-Experts models carry a growing share of frontier LLM serving, and
expert parallelism is how they deploy. The experts of each MoE layer spread across a
group of $G$ GPUs. Every token's hidden state travels to the ranks (the group's
member GPUs) that hold its experts, and the layer completes when the most loaded rank finishes~\cite{deepseek-aiDeepSeekV3TechnicalReport2025, yangQwen3TechnicalReport2025, lepikhinGShardScalingGiant2020}. The group therefore runs at the pace of its worst straggler,
once per layer, at every decode step.

\begin{figure}[t]
 \centering
 \includegraphics[width=0.92\linewidth]{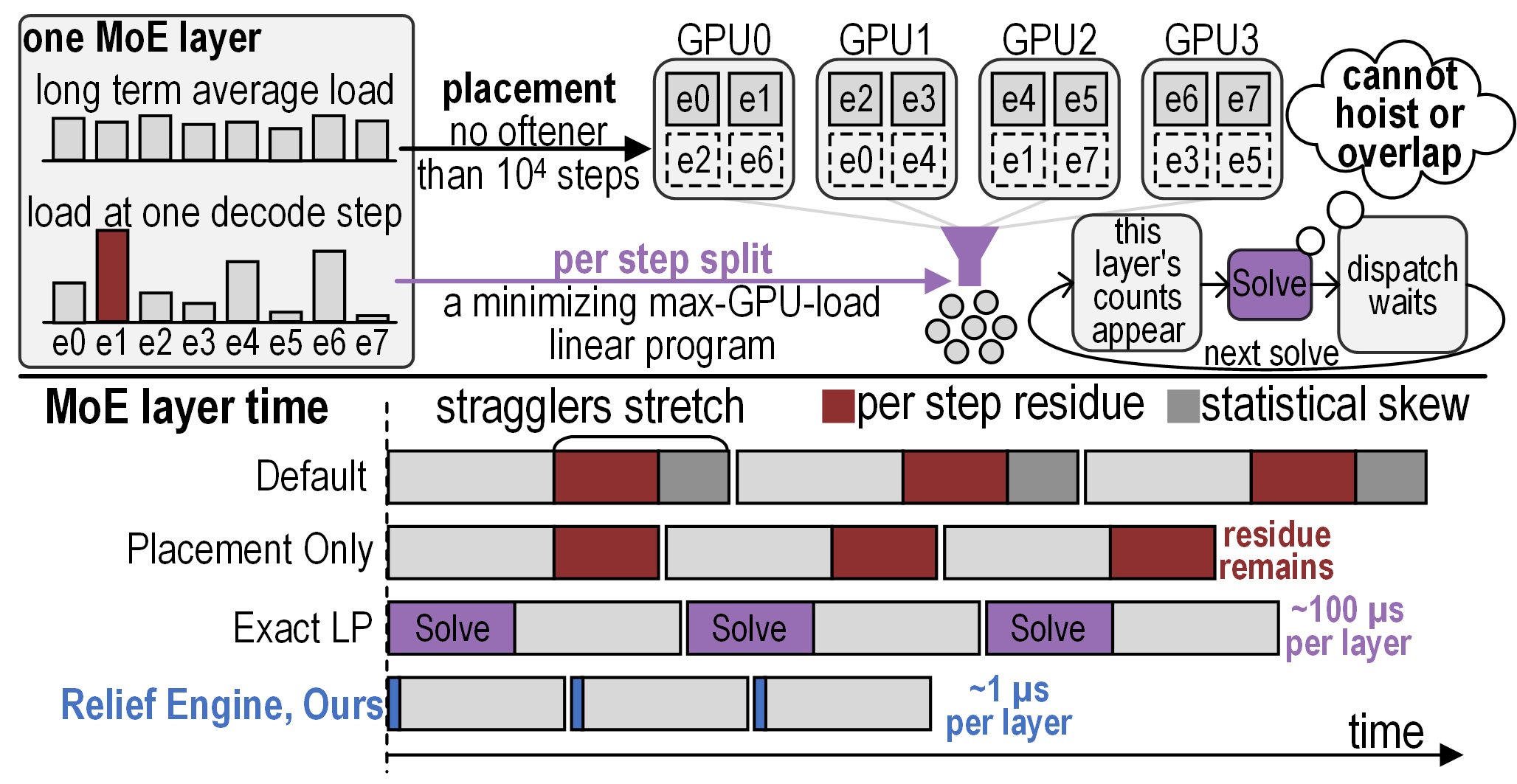}
 \caption{The per-step balancing problem. Expert load divides into a slow
 statistical skew, which periodic re-placement absorbs, and a per-step
 residue, visible only inside the decode iteration, which no deployed
 mechanism follows. The exact solver follows it and adds
 $\sim$100\,\textmu s to every layer, and the bottom row follows it at
 this paper's measured silicon latency. Each block of the lower timeline
 is one MoE layer, and a decode step runs every layer.}
 \label{fig:landscape}
\end{figure}

Token-to-expert assignments concentrate on few experts and shift from step
to step. The ranks therefore finish unevenly, and the group's reserved GPU time
waits on whichever rank drew the most tokens. The waste is reported wherever
MoE serving is measured. Published characterizations place MoE-layer compute
utilization below 11\% on GPUs~\cite{yunDuplexDeviceLarge2024}, while the
decode phase draws roughly half of serving's GPU
power~\cite{patelSplitwiseEfficientGenerative2024a}. On decode routing captured
from a 128-expert open model, an unbalanced 64-GPU group performs
expert work during a quarter of the GPU time it reserves
(\S\ref{sec:model}).

Deployed systems balance today with one static mechanism
(Fig.~\ref{fig:landscape}). EPLB
periodically recomputes which experts are replicated on which GPUs, a placement
decision that migrates expert weights and therefore runs minutes
apart~\cite{DeepseekaiEPLB2026}. Between placements the production engines pick among
an expert's replicas by position or at random, oblivious to load. An exact
per-step mechanism also exists. LPLB re-solves a min--max linear program
over the replica split at every batch on the GPU~\cite{DeepseekaiLPLB2026}.
Neither mechanism recovers the idle GPU time. The static mechanism cannot track per-step routing, and
the distance it leaves to the per-step optimum does not close at any
scale
(Fig.~\ref{fig:staircase}). The exact mechanism fails on its own latency. The counts a per-step split
consumes are produced inside the very iteration it serves, so the decision
cannot leave the decode loop. The layer dispatches only after the split
arrives. For layers near its own $\sim$100\,\textmu s solve
latency the decision therefore costs more utilization than its balance
recovers. Further, its factorization's working set grows with the square of the
group against a fixed on-chip capacity, so at deployment scale it stops
launching at all (\S\ref{sec:model:premise}). Both failures have been read as fundamental.
A recent ISCA paper asserts that MoE's dynamism prevents ``any
load-balancing solutions to better distribute the number of activated
experts across the GPUs in an even manner''~\cite{hwangPregatedMoEAlgorithmSystem2024}.

The decision-frequency measurement of \S\ref{sec:model:premise} contradicts
that reading. Reusing a balancing split for even one additional decode step
forfeits 62--91.6\% of the utilization it recovers over the static baseline,
and a free, clairvoyant periodic re-placement collapses to the static
baseline as well. No schedule coarser than one decision per step therefore realizes the value
a per-step decision holds. That optimum is set by the step's own counts,
and the exact solvers reach it at a latency the layer cannot absorb. Reaching that balance does not require the
program. Relief is a bidirectional pairwise relaxation over the placement's
replica edges, in integer arithmetic throughout, and a few passes of
halving transfers bring the load table to within a fraction of a point of
the per-step optimum. Moreover, our best GPU kernels of it already outrun
the exact solver, 3.7 to 7.3$\times$ call for call, and keep running where
the solver cannot launch (\S\ref{sec:deploy:kernel}). What latency remains sits in
storage. Every pass re-reads one small table at indices known only at run
time, and which storage can hold that table is what separates the
deployments of this method (\S\ref{sec:engine}).

We therefore treat deployment as part of the design and build the Relief solve three ways. The kernel runs on one of the serving accelerator's own SMs. The same
verified engine runs on an FPGA across PCIe, field-reprogrammable as the deployment
evolves. A small fixed-function block sits beside the serving datapath, one per
accelerator, of the kind Intel ships for load balancing in Xeon
processors~\cite{louDynamicLoadBalancer2025a}. Every deployment consumes the raw per-expert counts in 32-bit fixed point
and returns a bit-identical split.

On decode routing from three open MoE families served with
expert parallelism at 8 to 64 GPUs, scored as the share of the group's
reserved GPU time that performs expert work relative to the best split any
policy could produce each step, the deployed placement scheme realizes 66
to 72\% of that bound by model family. Our kernel raises the three
families to 93\%, and our silicon deployment raises them to 99.5\%, while adding
0.02\% of a modern accelerator's transistors. The largest recovery over the deployed scheme is 37.3 points of that
share. At $G{=}64$ a fixed ratio on the fitted placement delivers
less utilization than no balancing at all, a reversal
\S\ref{sec:model:premise} traces to the placement's own structure.

This paper makes three contributions.
\begin{itemize}
\item \textbf{Relief, an all-integer split algorithm that makes the exact
solver unnecessary.} A few passes of bidirectional pairwise halving along the placement's
replica edges come within a fraction of a point of the per-step optimum,
and the working set is the load vector itself.
\item \textbf{We measure three deployments of the method, from software to
silicon.} The kernel outruns the exact solver on the solver's own GPU, and
the same verified engine holds the load table in registers on an FPGA across
PCIe and as $0.14$\,mm$^2$ of silicon beside the accelerator. No deployment
is a prerequisite of the method, and the evaluation compares all three
under one metric.
\item \textbf{Effective utilization scores balance jointly with decision
latency.} A split loses most of its gain after one step of reuse, so a
scheme is only as good as the decision it can repeat every step. We charge
each scheme its decision latency against the MoE layer it serves and sweep
that layer's duration, covering every interconnect, batch, and GPU
generation in one study.
\end{itemize}

\section{Motivation and the Model}
\label{sec:model}

\subsection{Expert imbalance leaves provisioned GPU time idle}
\label{sec:model:imbalance}

\begin{figure}[t]
 \centering
 \includegraphics[width=\linewidth]{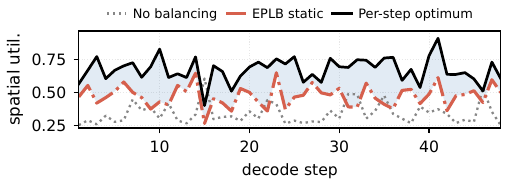}
 \caption{Per-decode-step spatial utilization, layer 24, $G{=}16$.
 Routing moves every step. A placement fitted to the layer's mean (EPLB
 static) lifts the trace but tracks none of the motion, while a split
 re-solved on each step's own counts follows it. The shaded band is what a
 per-step decision recovers. The optimum stays near
 0.67, which is the balance this fixed two-replica placement admits. The
 remaining distance to 1.0 is set by the placement, which no split can
 move.}
 \label{fig:steps}
\end{figure}

An expert-parallel MoE layer distributes its experts across a group of $G$ GPUs and
completes each decode step at the pace of the most loaded rank. Routing concentrates
tokens on few experts, so the ranks finish unevenly. The gap between the mean and
the maximum per-rank load is GPU time the deployment provisions and cannot use. On
real routing from a 128-expert model (Qwen3-30B-A3B), the spatial utilization of an
unbalanced group, the ratio of mean to maximum per-rank load, averages 0.42 at
$G{=}16$ and falls to 0.24 at $G{=}64$ across the model's 48 layers. At deployment
scale, three quarters of the provisioned GPU time in the expert stage sits
idle behind stragglers, in the regime where fewer than twenty tokens are
in flight for 60--70\% of machine
time~\cite{patelSplitwiseEfficientGenerative2024a}. Figure~\ref{fig:steps}
records the volatility itself. On one layer at $G{=}16$, routing shifts every
decode step. A split frozen from the layer's mean tracks none of the motion, while
a split re-solved each step holds the per-step optimum throughout.

In the serving engines that carry production traffic, the per-token choice among
an expert's replicas, the \emph{replica pick}, consults no load signal. vLLM resolves a replicated
expert by the token position's
modulo~\cite{kwonEfficientMemoryManagement2023b}, and SGLang draws the
replica uniformly at random~\cite{zhengSGLangEfficientExecution2024}. The balancing that deployments do
perform operates on a different mechanism and a different timescale. EPLB periodically
recomputes which experts are replicated on which GPUs~\cite{DeepseekaiEPLB2026}, a decision
that migrates expert weights between devices and therefore runs no oftener
than minutes apart. DeepSeek adjusts its redundant experts every ten
minutes when prefilling and states no interval for
decoding~\cite{deepseek-aiDeepSeekV3TechnicalReport2025}, so thousands of
decode steps separate two placements. Between placements the deployed
systems hold the split they were given, and the per-step motion of
Fig.~\ref{fig:steps} passes unanswered.

\subsection{The two balancing mechanisms and their timescales}
\label{sec:model:levers}

MoE load balancing has two mechanisms, and they differ in what moving them costs.
The \emph{placement} mechanism decides which experts, and how many replicas of each, live
on which GPU, and moving it copies expert weights across devices. The \emph{split} mechanism
decides, for a fixed placement, what fraction of a replicated expert's tokens each
replica receives, and moving it rewrites a routing table and moves no weights at all.
The imbalance itself divides the same way. One part is a slow statistical
skew, the experts that are hot on average, which is what placement absorbs
and why a weight-migrating decision need run no oftener than minutes apart,
if it runs again at all. The other part is a per-step residue, the deviation
of each step's counts from that average, readable only in the step itself.
The gating that produces those
counts runs inside the decode iteration they serve, so whatever tracks the
residue must decide inside that iteration, between the routing that produces
the counts and the dispatch that waits on the split
(Fig.~\ref{fig:landscape}).

The replicated experts form a graph whose edges join the two ranks that
hold a copy, and the split chooses one fraction per edge
(Fig.~\ref{fig:landscape}, top). Everything upstream of the split is
already deployed practice, and what the step's counts leave open is the
fraction on each edge.

EPLB is deployed and re-decides only the placement. A recent line of exact
solvers re-decides the split at every step in software. LPLB re-solves a
min--max linear program over the split ratios on the GPU at every
batch~\cite{DeepseekaiLPLB2026}. No deployed mechanism
re-decides the split at every step at a latency the step can absorb.
\S\ref{sec:model:premise} traces that absence to its cost, and
Table~\ref{tab:tiers} summarizes the mechanisms.

\begin{table}[t]
\centering\scriptsize
\caption{Balancing mechanisms compared by update rate, use of the current
step's loads, per-step decision latency, availability at $G{=}64$, and
whether the mechanism can reach the per-step optimum at all.
\S\ref{sec:model:premise} quantifies the latency and the launch limit.}
\label{tab:tiers}
\setlength{\tabcolsep}{2.4pt}
\begin{tabular}{@{}lccccc@{}}
\toprule
Scheme & Rate & Load-aware & Latency & $G{=}64$ & Optimum \\
\midrule
Replica pick (deployed) & per step & $\times$ & $\sim$0 & \checkmark & $\times$ \\
EPLB placement (deployed) & minutes & $\times$ & 0$^{a}$ & \checkmark & $\times$ \\
Exact split, software & per step & \checkmark & 102\,\textmu s & $\times$ & \checkmark \\
Relief (ours, three deployments) & per step & \checkmark & 0.75--28\,\textmu s & \checkmark & \checkmark$^{b}$ \\
\bottomrule
\end{tabular}

\smallskip
{\scriptsize $^{a}$weight migration charged at zero throughout
(\S\ref{sec:eval:method}). $^{b}$to within a fraction of a point; the fixed point every deployment
runs attains 99.5 to 99.8\% of it (Table~\ref{tab:ablation}).}
\end{table}

\subsection{Effective utilization and the per-step requirement}
\label{sec:model:ueff}

\begin{figure*}[t]
 \centering
 \includegraphics[width=\textwidth]{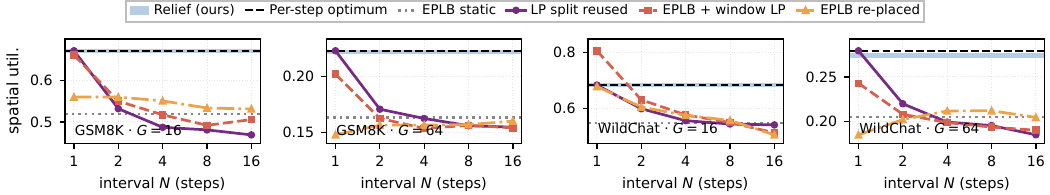}
 \caption{The per-decision split against periodic balancing on layer 24, both
traces and both group sizes, with every periodic scheme's re-solve charged
at zero and each window's own future statistics granted. The Relief band coincides with the per-step
 optimum in every panel, and every periodic scheme leaves it. The one
 oracle exception (panel c), clairvoyant joint re-fitting at $N{=}1$,
 migrates expert weights every decode step.}
 \label{fig:cadence}
\end{figure*}

The metric follows from GPU-time accounting over one MoE layer at one
decode step, the pair every per-step quantity in this paper names. The step's routing scores each token onto $k$ of the layer's $E$ experts
and leaves per-expert token counts $n_1,\dots,n_E$; the placement maps each
expert copy to one of the $G$ ranks; and the split divides a replicated
expert's count between its two copies. Together the three turn the $E$
per-expert counts into per-rank loads $L_1,\dots,L_G$, the tokens each GPU
serves this step (Eq.~\ref{eq:lp} below gives $L_i$ in closed form). Each rank
then computes for a time proportional to its own load. The layer's
synchronous collectives nevertheless hold the whole group until the most loaded rank
finishes, so the stage reserves $G \cdot \max_i L_i$ of rank-time to perform
$\sum_i L_i$ of expert work. The decision adds latency ahead of the stage. Its input is the counts the
iteration just produced, and its output is what the dispatch waits on. It can
therefore neither leave the loop nor start early. A solve of latency \tsolve{} holds
all $G$ ranks that much longer while doing no expert work. Effective
utilization is the busy share of everything the group reserves, the product of
the balance the split achieves and the share of elapsed time not consumed by
the decision,
\begin{equation}
 \ueff =
 \underbrace{\frac{\sum_{i=1}^{G} L_i}{G\,\max_i L_i}}_{\text{balance: }\mathrm{spatial}}
 \cdot
 \underbrace{\frac{\tmoe}{\tsolve + \tmoe}}_{\text{share of time executing}}
 = \frac{\mathrm{spatial}}{1 + \tsolve/\tmoe},
 \label{eq:ueff}
\end{equation}
where \tmoe{} is the per-GPU time of everything the layer does once the
split is known, the dispatch all-to-all, the expert computation, and the
combine all-to-all. The first factor, hereafter $\mathrm{spatial}$, is the
mean-to-maximum ratio of per-rank load, and the paper carries the
right-hand form. Integrated over a decode trace at a fixed operating point, total useful over
total reserved GPU-time is the same expression with the trace's mean
$\mathrm{spatial}$ in the numerator, and every number in \S\ref{sec:evaluation}
is computed exactly so. The paper ranks schemes by this joint
quantity.

Every property of a deployment that a balancing scheme does not control enters
\ueff{} by moving \tmoe{}. The layer time decomposes as
$\tmoe = T_{\mathrm{disp}} + T_{\mathrm{exp}} + T_{\mathrm{comb}}$. Each all-to-all takes a fixed latency term plus its payload over the link
bandwidth, $T_{\mathrm{disp}} \approx \alpha + B\,k\,d_{h}\,b / (G\,W)$ for
batch $B$, top-$k$ routing, hidden width $d_{h}$, and $b$-byte
activations. The expert computation between them scales as $B/G$ on the
local GEMM rate. Every such setting enters through one term, so sweeping
\tmoe{} end to end covers every combination at once. The sweep follows
Turbo's methodology~\cite{seyedroudbariTurboSmartNICenabledDynamic2023},
and it lets a single-GPU study parameterize a 64-GPU deployment with every
group-size-dependent quantity measured at its own $G$. To place deployments on the
range, DeepEP's low-latency decode kernels
report the two collective terms alone at 313\,\textmu s for 16 ranks and
487\,\textmu s for 64, over CX7 RDMA at $W \approx
47$\,GB/s~\cite{DeepseekaiDeepEP2026}, before a single expert multiplies, so
production decode sits in the upper half of the swept range. A
prefill-scale batch raises \tmoe{} until any decision latency becomes
negligible, and its thousands of tokens smooth the per-step routing itself.
The contested region, and this study's traces, are therefore decode. The
accounting is conservative, since every scheme's decision is charged as
fully blocking. Moreover, the periodic alternatives of \S\ref{sec:model:premise} are
granted their re-optimization at zero cost, and the split's own
communication, a routing table and no expert weights, is inside
\tsolve{}.

The reference for the numerator is the \emph{per-step optimum}, the best split any
policy could produce on the given placement at the given step. It is the
solution of an explicit program, the instance of Fig.~\ref{fig:landscape} written
in symbols. On a group of $G$ GPUs, rank $i$ carries a fixed
load $f_i$ from its non-replicated experts and $D$ replicated slots, two
throughout the main study. Slot
$(i,d)$ has load $w_{i,d}$ and a second copy on the peer $\sigma(i,d)$ the placement
assigns, and $x_{i,d}\in[0,1]$ is the fraction of $w_{i,d}$ that stays on rank $i$.
Rank $i$'s load is then
$L_i(x)=f_i+\sum_{d} w_{i,d}\,x_{i,d}+\sum_{(j,d'):\,\sigma(j,d')=i} w_{j,d'}(1-x_{j,d'})$,
the share it keeps plus the share its peers offload onto it, and balancing the group
minimizes the peak:
\begin{equation}
 \min_{x,\,z}\; z \quad\text{s.t.}\quad L_i(x)\le z\ \ \forall i,\qquad
 0\le x_{i,d}\le 1.
 \label{eq:lp}
\end{equation}
Total load is conserved across any split, so minimizing the peak $z$ maximizes the
mean-to-peak ratio, which is the $\mathrm{spatial}$ term of Eq.~\ref{eq:ueff}. The
program optimizes the very numerator this paper scores against. The exact
solvers solve Eq.~\ref{eq:lp} directly, LPLB with an interior-point method
on the
GPU~\cite{DeepseekaiLPLB2026} and FineMoE on a
CPU~\cite{zhaoFinegrainedMoELoad2026}. The paper takes its solution as the per-step
optimum at zero cost and measures every scheme as a distance to this bound.
Under Eq.~\ref{eq:ueff} the schemes of \S\ref{sec:model:levers} differ
only in which factor they give up, as Table~\ref{tab:tiers} sets out. What
remains is to establish at what rate the decision must be made, and what
making it costs in software.

\label{sec:model:premise}%
Three measurements establish what the missing per-step mechanism requires.

\textbf{One step of reuse forfeits 62--91.6\% of a split's gain.} Re-solving every $N$ steps and reusing the split in between avoids the
per-step latency. We grant that
strategy its best case, a solver re-run on each window's own mean load with the
re-solve charged at zero cost, and measure it on real routing
(Fig.~\ref{fig:cadence}). The reused split falls below the static baseline it was meant to beat by
$N{=}4$ in three of the four trace and group-size settings, and by $N{=}8$
in all four. A fitted split tracks the step it was fitted to, while the next step's
routing already differs. The same experiment rules out the
other alternative. Granted free and clairvoyant re-placement at any interval, with a
re-fitted frozen split, the periodic scheme collapses to the static
baseline by $N{=}2$ at $G{=}64$ on GSM8K and by $N{=}4$ in two of the
three remaining settings. The last, WildChat at $G{=}16$, crosses only
at $N{=}16$ and still keeps 61.4\% of the per-step gain at $N{=}2$. Weight migration limits the
placement to statistics gathered over long windows, and the per-step
deviation is visible only to the split. The same limit reaches any placement fitted from window statistics,
popularity-proportional replication
included~\cite{nieFlexMoEScalingLargescale2023}, since the per-step
deviation is not in those statistics. On GSM8K at $G{=}64$ the per-step split wins all 48 layer
records at every $N{\geq}2$. The missing mechanism must therefore
decide once per layer per decode step, and cheaply.

\textbf{At that rate, the software solve costs more utilization than its balance recovers.} LPLB's
solver takes roughly 100\,\textmu s per solve on the GPU, a figure its own
documentation reports and our measurement confirms at
102\,\textmu s. Eq.~\ref{eq:ueff} turns that latency into a
utilization loss larger than the balance gain across the range's lower reach.
Below $\tmoe \approx 103$\,\textmu s at $G{=}16$, the exact solver's effective
utilization falls under the no-balancing baseline.

\textbf{At deployment scale the exact GPU solver no longer launches.} Equation~\ref{eq:lp} has
$\mathrm{NC}{=}G{+}GD$ load and box constraints for $D$ replicated slots per rank,
$3G$ at the two slots per rank the placements use. An interior-point solver
factors their $\mathrm{NC}{\times}\mathrm{NC}$ normal-equations matrix once per
iteration, and its Cholesky factor occupies $O(\mathrm{NC}^2)$ shared memory, a footprint
fixed at compile time by $G$ alone. The solver kernel's own allocation gives it in closed form,
$156G^2 + 160G + 40$ bytes on the $D{=}2$ topology family, and that
expression reproduces every measured launch footprint exactly. The GPU we measure admits 99\,KiB of dynamic shared memory per thread
block after opt-in, and the solver stops launching at $G{=}25$, the first
size whose footprint exceeds that capacity and the point a launch sweep
confirms. The same footprint fits $G{=}32$ inside an A100's 163\,KiB and
inside the 227\,KiB of H100 and Blackwell datacenter parts, so the group
size at which the solver stops launching follows the part. At $G{=}64$ the
footprint reaches 634\,KiB and exceeds the
per-thread-block capacity of every current architecture, by $2.8\times$ even
against the 227\,KiB that H100 and Blackwell datacenter parts offer, the
largest opt-in allocation any shipping architecture
provides~\cite{CUDAProgrammingGuide}.

\begin{figure}[t]
 \centering
 \includegraphics[width=\linewidth]{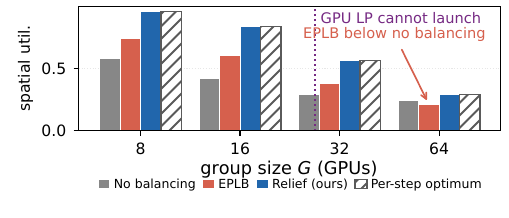}
 \caption{Spatial utilization by group size on real 128-expert GSM8K routing
 (layer-major, means of per-layer results), for no balancing, EPLB's static split, the Relief
 split, and the per-step optimum (hatched). Past the dotted rule the exact GPU solver does not launch on the measured GPU, and at $G{=}64$ the static split falls below no
 balancing. \S\ref{sec:eval:headline} gives the margins these bars
 show.}
 \label{fig:staircase}
\end{figure}

 Figure~\ref{fig:staircase} shows what
remains at scale in cost-free balance alone. At $G{=}64$ the fitted
placement's fixed ratios measure 0.21 spatial utilization, while the same replica
slots under the unfitted default placement measure 0.24, with re-placement
and weight migration charged at zero. The per-step optimum, solved on
the fitted placement, reaches 0.29. The fitted placement falls below the
default because its duplicate map presumes a split that adapts. The planner
spreads experts across ranks by their window means and copies each rank's
heaviest expert to a fixed neighbour, so the copies of a layer's two
heaviest experts share one rank (in every $G{=}64$ layer we probed). Under
the deployed replica pick that rank's realized load runs a median
$6.8\times$ the median rank's. The same planner fitted to a randomly
permuted mean, a placement of identical structure carrying no information
about the experts, measures 0.24 and matches the default, which isolates
the fit as the source of the loss.
Solving the fractions each step on the same placement reaches the 0.29,
so the fitted placement helps only where a per-step mechanism exists. Without one the deployment does better leaving the default
placement in place.

The missing mechanism must therefore return the per-step split at a
latency negligible against \tmoe{}, at group sizes where the exact GPU
solver does not launch. The sections that follow build it.

\section{The Relief Algorithm}
\label{sec:algorithm}

\begin{figure*}[t]
 \centering
 \includegraphics[width=0.82\linewidth]{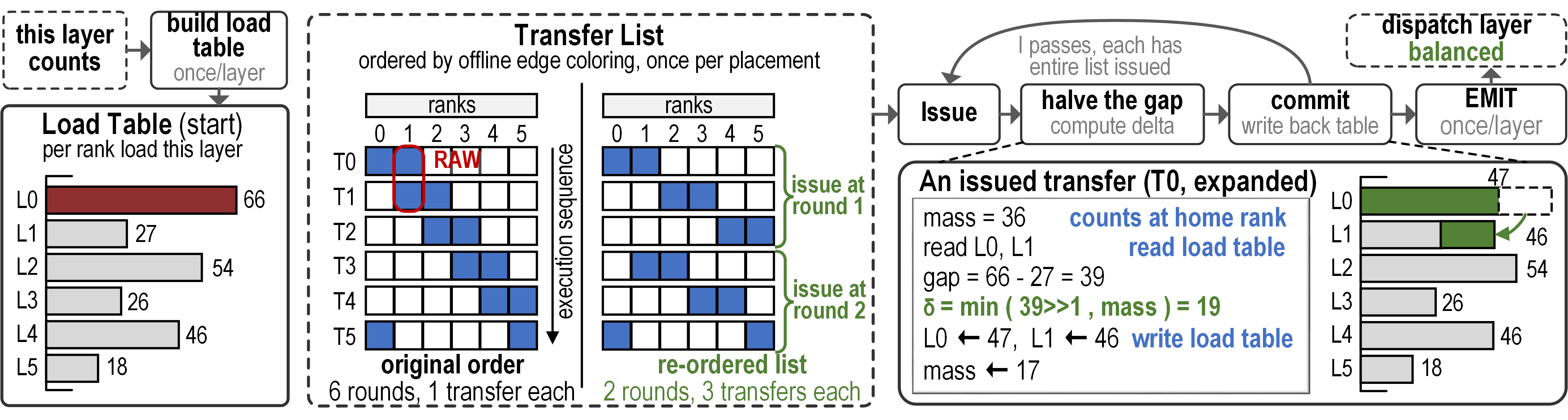}
 \caption{From counts to emitted fractions. In list order (left matrix)
 every consecutive pair of the six transfers shares a rank and needs its
 own round, while the placement-time colouring (right matrix) packs the
 same six into two rank-disjoint rounds of three. The expanded transfer
 shows one issue exactly: gap $66-27=39$,
 $\mathit{delta}=\min(39{\gg}1,\mathrm{mass})=19$, writebacks 47 and 46, the truncated halve leaving the two ranks one
 apart.}
 \label{fig:instrstream}
\end{figure*}

Balancing on a fixed placement leaves one quantity free. Each replicated
expert's tokens divide between its two copies, and that division is one
fraction per edge of the replica graph of \S\ref{sec:model:levers}. Every
choice of fractions is scored by the same live object, the table of
per-rank loads for the current step, and flattening this table is the whole
objective. The exact solvers reach the fractions through the min--max
program of Eq.~\ref{eq:lp}~\cite{DeepseekaiLPLB2026,
zhaoFinegrainedMoELoad2026}. Relief edits the table directly (Fig.~\ref{fig:instrstream}).

A pass visits each transfer in turn and applies the move that is locally
most effective for its edge. The move shifts half of the load gap between
the transfer's two ranks, and it is truncated to the tokens still available
on the sending side. Halving is the amount that levels the two ranks
exactly. The cap keeps every fraction inside its range. Because the sign of
the gap sets the direction, load flows toward the lighter copy from either
side of the edge.

Each move lands in the load table immediately, and the next transfer reads
the updated loads. The transfers therefore acquire an order, and a pass
becomes a walk of a transfer list against the live table. The convergence
depends on that live read, because when every transfer is instead computed
from a snapshot taken at the start of the pass, the same balance arrives
only after four times the passes (\S\ref{sec:eval:ablation}). A configured
number of passes $I$ repeats the walk. Algorithm~\ref{alg:relief} states
one pass, and the procedure is a Gauss--Seidel relaxation of
Eq.~\ref{eq:lp} in integer arithmetic.

\begin{algorithm}[t]
\caption{One Relief pass, and the emission that follows the last. $L$ is
the $G$-entry load table of Eq.~\ref{eq:lp};
for a transfer $t$ between the two ranks that hold an expert's copies, its home
$u$ and its host $v$, $\mathrm{tok}[t]$ is
that expert's token count this step and $\mathrm{mass}[t]$ how many of them are still
at $u$. Every quantity is an integer, and $\epsilon$ is one least significant bit.}
\label{alg:relief}
\begin{algorithmic}[1]
\FOR{each round $R$ of the schedule}
 \FOR{each transfer $t=(u,v) \in R$, together}
 \STATE $\mathit{gap} \gets L[u] - L[v]$
 \STATE $\mathit{delta} \gets |\mathit{gap}| \gg 1$
 \COMMENT{truncating halve}
 \STATE $\mathit{limit} \gets \mathit{gap}>0 \;?\; \mathrm{mass}[t] \;:\; \mathrm{tok}[t]-\mathrm{mass}[t]$
 \COMMENT{still at home, or already sent}
 \STATE $\mathit{delta} \gets \min(\mathit{delta},\, \mathit{limit})$
 \IF{$\mathit{delta} > \epsilon$}
 \STATE $L[u] \mathrel{-}= \mathrm{sgn}(\mathit{gap})\,\mathit{delta}$; \;
 $L[v] \mathrel{+}= \mathrm{sgn}(\mathit{gap})\,\mathit{delta}$; \;
 $\mathrm{mass}[t] \mathrel{-}= \mathrm{sgn}(\mathit{gap})\,\mathit{delta}$
 \ENDIF
 \ENDFOR
 \STATE commit the round; the next round reads what it wrote
\ENDFOR
\STATE after $I$ passes: $\mathrm{o\_weight}[t] \gets \mathrm{mass}[t] / \mathrm{tok}[t]$,
one reciprocal per replicated expert
\end{algorithmic}
\end{algorithm}

A pass walks the transfers one at a time because consecutive transfers may
touch the same rank. When they do, the later transfer must read the load
written by the earlier one, and any reordering changes the balance the pass
reaches. Two transfers that share no rank, however, read and write disjoint
entries of the table, so issuing them together produces the same table as
issuing them one after the other. The parallelism is therefore discovered
inside the serial order, in the way a scheduler finds independent
instructions inside a serial program.

Whether two transfers conflict depends only on the ranks named in their
edges, and the placement fixes those ranks. The schedule therefore needs no
count from any step. Software computes it once, when the placement changes,
as a proper edge colouring of the replica graph. Every solve until the next
placement walks the same compiled list. The graph has $G$ nodes and $GD$
edges, and its colouring is recomputed only when weight migration moves the
placement, minutes apart (\S\ref{sec:model:levers}). Every rank
touches $2D$ edges, so $2D$ rounds is a lower bound that greedy colouring
reaches on every configuration we measure. Further, a
$2D$-regular graph coloured in $2D$ colours makes every colour class a
perfect matching. Each round therefore holds exactly $\lfloor G/2\rfloor$
transfers, and the classes tile the list evenly, which is the property
behind the fixed-stride issue of \S\ref{sec:engine:uarch2}. The placement
hands the solve one list of $GD$ entries in colouring order. Each entry
names the two ranks of its transfer and carries no round field, since
position alone determines the round.

The last choice inside the algorithm is how many passes to run. Each pass
improves the balance, and each pass lengthens \tsolve{}, so \ueff{} is a
function of the pass count. Substituting the pass count into Eq.~\ref{eq:ueff} and solving for the
maximizer states the choice in closed form,
\begin{equation}
 I^{*} = \operatorname*{arg\,max}_{p \le p_{\max}}\;
 \frac{s[p]}{1 + \big(L_{\mathrm{if}} + (a + b\,p)\,P\big)/\tmoe},
 \label{eq:istar}
\end{equation}
where $s[p]$ is the balance the relaxation reaches in $p$ passes, measured
once for a placement family, and the second term of the denominator is the
deployment's own \tsolve{} at $p$ passes, from its interface latency
$L_{\mathrm{if}}$, its cycle model $(a,b)$, and its clock period $P$. The
scan is short, because the balance gained per pass saturates
(\S\ref{sec:eval:dial}). Further, it cannot run inside the solve, because
it needs \tmoe{}, the duration of the layer being served, and only the
driver knows that duration. Balance itself never indicates when to stop, since it improves with every
pass. After every pass the load table gives the balance, so one
instrumented solve records $s[p]$ for every $p$ at once, and a few such
solves at a new placement give the whole curve. The driver
therefore evaluates the loop per group size and layer duration once,
offline, and writes the result into each call as a plain field, a kernel
argument or a descriptor word.

How near the relaxation comes to the per-step optimum is measured in
\S\ref{sec:eval:ablation}, in both arithmetics. The fixed-point form
every deployment
runs is integer addition, comparison, and shift, with one reciprocal at
emission. A pass performs $G{\cdot}d$ work in a number of rounds the
placement's degree fixes, so group size reaches the algorithm only through
the pass count. \S\ref{sec:eval:dial} measures how that count moves with
scale, against an interior-point working set that grows as $G^2$.

\section{Three Deployments of the Relief Solve}
\label{sec:engine}

\begin{figure*}[t]
 \centering
 \includegraphics[width=0.93\textwidth]{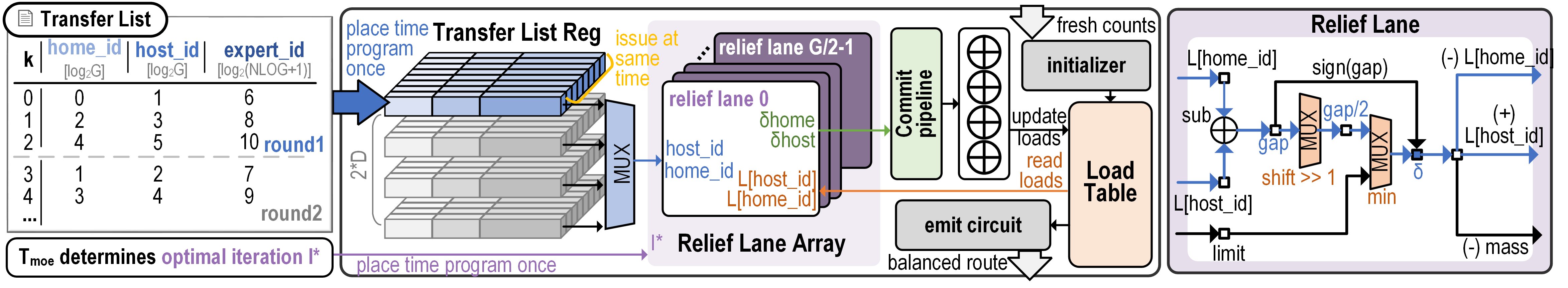}
 \caption{The engine. The load table is the datapath's only run-time index, and
 every round both reads and updates it. The transfer list and the pass
 count $I^{*}$ (Eq.~\ref{eq:istar}) arrive over the same
 program-once path. The right panel expands one lane, replicated $G/2$
 times.}
 \label{fig:fblock}
\end{figure*}

Relief is a fixed-point integer algorithm whose whole working state is one
$G$-entry table and one compiled list, and computation of that shape runs
on hardware at every level of a serving stack. This section implements it
three ways and measures each one. It runs as a CUDA kernel on the serving
GPU, as a silicon block beside the accelerator, and as an FPGA behind
PCIe. All three run the same relaxation on the same schedule and return
bit-identical splits. They differ in the storage that holds the load table,
and therefore in the time a solve takes.

Every deployment leaves the two mechanisms of \S\ref{sec:model:levers} on
their own clocks, the separation the decision-frequency experiment measured
(\S\ref{sec:model:premise}). The serving stack keeps the placement it already
has, re-fitted at EPLB's own interval, and the split is applied every step on
top of it.

\subsection{The Kernel}
\label{sec:deploy:kernel}

The kernel runs the whole solve inside one thread block. Shared memory
holds the load table and the mass each transfer still keeps at home, and
the compiled transfer list arrives as an argument. A pass steps through the
list one round at a time. Within a round each lane reads the two loads its
transfer names, halves the gap, caps the move by the mass at home, and
writes both loads back. The ranks of a round are disjoint, so these writes
never collide and the kernel needs no atomic operation. A single barrier
separates one round from the next, and it is the only synchronization a
pass performs.

Two properties of the algorithm shape this implementation. The round
assignment never runs on the device, because it depends on the placement
alone (\S\ref{sec:algorithm}), so the device never scores, sorts, or
schedules and the list it walks is data. The arithmetic is the engine's
own, because a version in the engine's Q11.20 word matches the RTL
reference bit for bit at no additional time, so a serving stack that runs
the kernel and one that installs the engine obtain the same split. Its own footprint is the $G$-word load table and the $E$-entry list, so
it launches at every group size, including those past the exact solver's
limit (\S\ref{sec:model:premise}). \S\ref{sec:eval:headline} scores what
it returns and how quickly.

The transfer list supplies the load table's indices only at run time, and
a GPU register file cannot be indexed that way, so the operands live in
shared memory. Every dependent step
waits on the load-to-use path of roughly 25\,ns, with no store-to-load
bypass, and caches do not shorten it, since
the table always hits and the dependent step then waits the hit latency
itself. The kernel needs only the serving GPU a deployment
already owns, running on one SM's issue stream.

Every dependent step is a read--modify--write of two load-table
entries at run-time indices, and a software implementation performs
almost no arithmetic besides it. Solve time therefore tracks the storage that holds
the table between dependent steps, and storage is a variable a deployment
can choose. The two hardware deployments that follow choose it.

\subsection{Silicon Beside the Accelerator}
\label{sec:engine:why}
\label{sec:engine:uarch2}

The second deployment integrates the relaxation as silicon on the
accelerator. Every dependent round of the relaxation reads two entries of the
$G$-word load table at indices that the transfer list names and writes them
back.
In the kernel that table sits in shared memory behind the 25\,ns
load-to-use path (\S\ref{sec:deploy:kernel}). In silicon the same
table is $G$ words of Q11.20 held in registers whose outputs are wires into
the arithmetic, so the dependent read takes a wire delay instead of a memory
access. The rest of the engine follows from the schedule
(Fig.~\ref{fig:fblock}). A lane array walks the transfer list, a mass state
tracks what each transfer has moved, a reciprocal pipeline emits the
fractions, and a descriptor interface connects the engine to the datapath.

The transfer list is the compiled list of \S\ref{sec:algorithm}, entries
in colouring order, each naming its \emph{home} rank, \emph{host} rank,
and \emph{expert}. A round therefore issues by position alone, every
lane's candidates are a compile-time stride of the list, and every
transfer of a round issues unconditionally. The one run-time index in the
datapath is the load-table read that each entry's rank fields name.

The load table's registers feed the lane array through wires, and a round
takes two cycles. The lanes read the two
loads each transfer names, halve the gap, and cap it by the transfer's
remaining mass (Fig.~\ref{fig:fblock}, right). Then the round commits, the per-rank deltas landing at fixed
indices. The next round reads what this one wrote, preserving the
relaxation's sequential semantics exactly. Mass is tracked as integer tokens,
so the loop body holds no division. Once per solve, an eight-lane emit phase
converts each transfer's remaining mass to a fraction with one
reciprocal, itself precomputed in a two-lane pipeline hidden under the passes.
The lane array is sized at $L{=}G/2$ because a colouring round is a
perfect matching of exactly $\lfloor G/2 \rfloor$ transfers, so one round
fills one issue and a wider array could not issue rounds any sooner. The
$E$-wide scan runs in $\lceil E/L \rceil$ chunks over the same lanes.

On the 128-expert family a solve costs $14{+}8I$ cycles at $G{=}8$ and
$28{+}8I$ at $G{=}64$, measured in RTL simulation on the real routing
traces. The slope is eight there because that family's topologies colour
to their degree lower bound of four rounds and a round costs two cycles,
so only the fixed overhead grows with $G$. Every family's engine is priced on its own measured counts
(\S\ref{sec:eval:method}), which Table~\ref{tab:verdict} charges row by
row, and each hardware deployment multiplies them by its own clock
period. The RTL is bit-exact against the fixed-point
reference on 128 of 128 verification programs, so every deployment of the
engine computes the same split the kernel does.

The engine sits beside the serving datapath as a bus master, one per
accelerator. It consumes the counts that the layer's gating just produced on
that die, and it emits the split that the dispatch waits on
(\S\ref{sec:model:premise}). Every rank routes by the same split, so a
centralized decision maker would add a broadcast to that window. LPLB
already avoids one by all-reducing the counts and solving the same instance
redundantly on every rank~\cite{DeepseekaiLPLB2026}, and the engine keeps
that arrangement. The split is therefore emitted where it is consumed, and
the interconnect carries only the counts routing was already
exchanging. One engine per die serves every layer of that die, because
the relaxation reads the $E$ per-expert counts and runs its measured cycle
count regardless of how many tokens the step carries. The
engine's duty cycle therefore depends on group size and step rate alone,
under 3\% of a layer's duration at every operating point of
Table~\ref{tab:verdict}.

\label{sec:engine:iface}
The bus interface is a descriptor. Software writes one word naming the
count buffer, the pass count, and the result buffer, and rings a
doorbell. The engine fetches the raw per-expert
counts and derives its internal per-replica view while they arrive, which
adds no cycles. It then relaxes and writes the split back. A solve issues only four bus transactions against a slave-port design's
385, a $96\times$ reduction. Charged at a conservative 0.43\,\textmu s for the platform's interface latency (the measured mastering
path itself takes 0.04\,\textmu s), the interface stays a minor term of
\tsolve{} at every operating point in \S\ref{sec:evaluation}. The
mastering logic adds 3.2\% to the engine's cell count in synthesis and sits
off its critical path.

Each round touches the load vector and the per-replica view, 768 bytes at
$G{=}64$, inside a register file of 4.6\,KiB in total
(Fig.~\ref{fig:fblock}).

At the clock-tree stage's 2.89\,ns period the relaxation takes
0.36\,\textmu s at $G{=}64$ and $I{=}12$. The descriptor interface adds 0.43\,\textmu s on top, so at this pass count the bus term
exceeds the relaxation at every group size and the interface sets \tsolve{}. The engine comes
to 0.14\,mm$^2$ of standard cells and ten million transistors
(\S\ref{sec:characterization}), and \S\ref{sec:eval:headline} scores it
against every other scheme under Eq.~\ref{eq:ueff}.

\begin{figure*}[t]
 \centering
 \includegraphics[width=\textwidth]{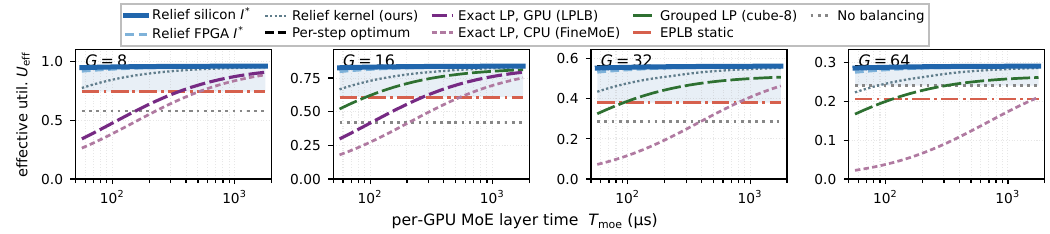}
 \caption{Effective utilization across the \tmoe{} range at every group
 size, layer-major. The shaded band is what the per-step split recovers
 over the best static scheme, and each panel carries only the schemes that
 run at that scale. Production decode layers measure
 313--487\,\textmu s in their collectives alone (\S\ref{sec:model:ueff}),
 in the range's upper half.}
 \label{fig:ueffpair}
\end{figure*}

\subsection{The FPGA Across PCIe}
\label{sec:deploy:fpga}

The engine can also be built as a PCIe device. A card added to the host
supplies the per-step split to a server already serving MoE models, with no
accelerator tape-out. The engine's dimensions are parameters of the RTL
(\S\ref{sec:engine:uarch2}), so the card is generated to the group size
the server runs. A placement change rewrites the transfer list, and an
algorithm change reloads the fabric under the same descriptor contract.
Implemented out of context for the XCVH1582 on a VHK158 card, the $G{=}64$
engine routes and closes timing at 7.2\,ns, or 139\,MHz. Every setup and
hold endpoint meets its constraint. The design occupies sixteen percent of
the device's LUTs
(Table~\ref{tab:overhead}). The arrangement differs from the silicon
deployment's in one respect. Where each accelerator's engine re-solves the
shared instance to keep the answer off the interconnect, the card already
sits across a bus. It therefore solves the group's instance once, and the host
distributes the returned split. The descriptor and the split cross PCIe,
half a kilobyte each way, and the solve is charged a conservative two bus
round trips, 2\,\textmu s against the literature's $\sim$0.9\,\textmu s
per small-payload round
trip~\cite{neugebauerUnderstandingPCIePerformance2018}. \tsolve{} comes to 2.9\,\textmu s at $G{=}64$ and $I{=}12$. The card
therefore lands between the two other deployments, because the same
relaxation runs on a slower fabric clock and crosses a bus that the on-die
engine does not. It returns the SM to serving tokens, and
\S\ref{sec:eval:headline} scores it beside the other two.

\section{Evaluation}
\label{sec:evaluation}

\begin{table*}[t]
\centering\footnotesize
\caption{Effective utilization across schemes, model families, and group
sizes. Each cell is \ueff{} as a percentage of the per-step optimum's, with
the scheme's measured \tsolve{} charged through Eq.~\ref{eq:ueff}. Columns
nest dataset, operating point (\S\ref{sec:eval:headline}), then $G$. Bold
marks the best value per column and the shortest \tsolve{} per panel. A cross marks a configuration that does not launch on the measured GPU,
and at $G{=}64$ one whose footprint exceeds every shipping architecture
(\S\ref{sec:model:premise}). A blank marks a configuration that does not
exist. $I^{*}$ rows re-solve the pass count per operating point. The exact-solver rows repeat across families, each granted the optimum's
balance and charged only its latency; a row's mean covers the cells it
has. $^{a}$the
same placement with the split solved exactly, on the
GPU~\cite{DeepseekaiLPLB2026} and on a CPU~\cite{zhaoFinegrainedMoELoad2026}.
$^{b}$the grouped solver solves $G/8$ independent eight-rank problems, so
both its latency and its balance ceiling belong to that smaller problem.
The kernel row's \tsolve{} is the latency its caller observes.}
\label{tab:verdict}
\setlength{\tabcolsep}{2.2pt}
\begin{tabular}{@{}lr*{9}{c}|*{9}{c}rr@{}}
\toprule
& & \multicolumn{9}{c}{GSM8K} & \multicolumn{9}{|c}{WildChat} & & \\
\cmidrule(lr){3-11}\cmidrule(lr){12-20}
\multicolumn{2}{@{}r}{$\tmoe$\,(\textmu s):} & \multicolumn{3}{c}{100} & \multicolumn{3}{c}{500} & \multicolumn{3}{c}{2000} & \multicolumn{3}{|c}{100} & \multicolumn{3}{c}{500} & \multicolumn{3}{c}{2000} & & \\
\cmidrule(lr){3-5}\cmidrule(lr){6-8}\cmidrule(lr){9-11}\cmidrule(lr){12-14}\cmidrule(lr){15-17}\cmidrule(lr){18-20}
Scheme & \multicolumn{1}{r}{$G$:} & 16 & 32 & 64 & 16 & 32 & 64 & 16 & 32 & 64 & 16 & 32 & 64 & 16 & 32 & 64 & 16 & 32 & 64 & \tsolve{} & mean \\
\midrule
\multicolumn{22}{@{}l}{\emph{Qwen3-30B-A3B, 128 experts}} \\
No balancing & & 50.3 & 51.0 & 82.8 & 50.3 & 51.0 & 82.8 & 50.3 & 51.0 & 82.8 & 59.6 & 57.0 & 77.6 & 59.6 & 57.0 & 77.6 & 59.6 & 57.0 & 77.6 &  & 63.0 \\
EPLB (deployed) & & 72.5 & 67.7 & 70.7 & 72.5 & 67.7 & 70.7 & 72.5 & 67.7 & 70.7 & 77.8 & 71.0 & 71.9 & 77.8 & 71.0 & 71.9 & 77.8 & 71.0 & 71.9 &  & 71.9 \\
LPLB (GPU LP)$^{a}$ & & 49.5 & $\times$ & $\times$ & 83.1 & $\times$ & $\times$ & 95.1 & $\times$ & $\times$ & 49.5 & $\times$ & $\times$ & 83.1 & $\times$ & $\times$ & 95.1 & $\times$ & $\times$ & 102 & 75.9 \\
FineMoE (CPU LP)$^{a}$ & & 32.6 & 20.8 & 12.9 & 70.7 & 56.8 & 42.5 & 90.6 & 84.0 & 74.7 & 32.6 & 20.8 & 12.9 & 70.7 & 56.8 & 42.5 & 90.6 & 84.0 & 74.7 & 207--676 & 54.0 \\
Grouped LP (cube-8)$^{b}$ & & 74.0 & 69.0 & 68.7 & 92.4 & 86.2 & 85.9 & 97.0 & 90.4 & 90.1 & 74.4 & 68.4 & 68.8 & 93.0 & 85.4 & 86.0 & 97.5 & 89.6 & 90.3 & 33$^{b}$ & 83.7 \\
EPLB + relief (kernel), $I^{*}$ & & 86.9 & 84.8 & 84.4 & 96.5 & 95.5 & 95.4 & 99.0 & 98.6 & 98.5 & 87.4 & 83.0 & 83.2 & 96.8 & 94.7 & 94.9 & 99.1 & 98.3 & 98.3 & 13--28 & 93.1 \\
EPLB + relief (FPGA), $I^{*}$ & & 97.1 & 96.7 & 96.7 & 99.3 & 99.2 & 99.2 & 99.8 & 99.8 & 99.8 & 97.2 & 96.5 & 96.5 & 99.4 & 99.2 & 99.2 & 99.8 & 99.8 & 99.8 & 2.8--4.1 & 98.6 \\
EPLB + relief (silicon), $I^{*}$ & & \textbf{99.1} & \textbf{98.9} & \textbf{98.9} & \textbf{99.8} & \textbf{99.7} & \textbf{99.7} & \textbf{99.9} & \textbf{99.9} & \textbf{99.9} & \textbf{99.2} & \textbf{98.8} & \textbf{98.8} & \textbf{99.8} & \textbf{99.7} & \textbf{99.7} & \textbf{99.9} & \textbf{99.9} & \textbf{99.9} & \textbf{0.7--1.7} & \textbf{99.5} \\
\midrule
& \multicolumn{1}{r}{$G$:} & \multicolumn{3}{c}{16\hspace{1.15em}32} & \multicolumn{3}{c}{16\hspace{1.15em}32} & \multicolumn{3}{c}{16\hspace{1.15em}32} & \multicolumn{3}{|c}{16\hspace{1.15em}32} & \multicolumn{3}{c}{16\hspace{1.15em}32} & \multicolumn{3}{c}{16\hspace{1.15em}32} & & \\
\cmidrule(l){1-22}
\multicolumn{22}{@{}l}{\emph{OLMoE-1B-7B, 64 experts}} \\
No balancing & & \multicolumn{3}{c}{58.3\hspace{1.15em}66.2} & \multicolumn{3}{c}{58.3\hspace{1.15em}66.2} & \multicolumn{3}{c}{58.3\hspace{1.15em}66.2} & \multicolumn{3}{|c}{58.3\hspace{1.15em}68.1} & \multicolumn{3}{c}{58.3\hspace{1.15em}68.1} & \multicolumn{3}{c}{58.3\hspace{1.15em}68.1} &  & 62.7 \\
EPLB (deployed) & & \multicolumn{3}{c}{68.8\hspace{1.15em}64.4} & \multicolumn{3}{c}{68.8\hspace{1.15em}64.4} & \multicolumn{3}{c}{68.8\hspace{1.15em}64.4} & \multicolumn{3}{|c}{72.6\hspace{1.15em}64.6} & \multicolumn{3}{c}{72.6\hspace{1.15em}64.6} & \multicolumn{3}{c}{72.6\hspace{1.15em}64.6} &  & 67.6 \\
LPLB (GPU LP)$^{a}$ & & \multicolumn{3}{c}{49.5\hspace{1.15em}$\times$} & \multicolumn{3}{c}{83.1\hspace{1.15em}$\times$} & \multicolumn{3}{c}{95.1\hspace{1.15em}$\times$} & \multicolumn{3}{|c}{49.5\hspace{1.15em}$\times$} & \multicolumn{3}{c}{83.1\hspace{1.15em}$\times$} & \multicolumn{3}{c}{95.1\hspace{1.15em}$\times$} & 102 & 75.9 \\
FineMoE (CPU LP)$^{a}$ & & \multicolumn{3}{c}{32.6\hspace{1.15em}20.8} & \multicolumn{3}{c}{70.7\hspace{1.15em}56.8} & \multicolumn{3}{c}{90.6\hspace{1.15em}84.0} & \multicolumn{3}{|c}{32.6\hspace{1.15em}20.8} & \multicolumn{3}{c}{70.7\hspace{1.15em}56.8} & \multicolumn{3}{c}{90.6\hspace{1.15em}84.0} & 207--380 & 59.3 \\
EPLB + relief (kernel), $I^{*}$ & & \multicolumn{3}{c}{86.7\hspace{1.15em}84.4} & \multicolumn{3}{c}{96.4\hspace{1.15em}95.2} & \multicolumn{3}{c}{99.0\hspace{1.15em}98.4} & \multicolumn{3}{|c}{87.1\hspace{1.15em}82.9} & \multicolumn{3}{c}{96.5\hspace{1.15em}94.9} & \multicolumn{3}{c}{99.1\hspace{1.15em}98.3} & 13--26 & 93.2 \\
EPLB + relief (FPGA), $I^{*}$ & & \multicolumn{3}{c}{97.0\hspace{1.15em}96.6} & \multicolumn{3}{c}{99.3\hspace{1.15em}99.1} & \multicolumn{3}{c}{99.8\hspace{1.15em}99.7} & \multicolumn{3}{|c}{97.2\hspace{1.15em}96.5} & \multicolumn{3}{c}{99.4\hspace{1.15em}99.2} & \multicolumn{3}{c}{99.8\hspace{1.15em}99.7} & 2.8--5.0 & 98.6 \\
EPLB + relief (silicon), $I^{*}$ & & \multicolumn{3}{c}{\textbf{99.0}\hspace{1.15em}\textbf{98.8}} & \multicolumn{3}{c}{\textbf{99.8}\hspace{1.15em}\textbf{99.7}} & \multicolumn{3}{c}{\textbf{99.9}\hspace{1.15em}\textbf{99.9}} & \multicolumn{3}{|c}{\textbf{99.2}\hspace{1.15em}\textbf{98.8}} & \multicolumn{3}{c}{\textbf{99.8}\hspace{1.15em}\textbf{99.7}} & \multicolumn{3}{c}{\textbf{99.9}\hspace{1.15em}\textbf{99.9}} & \textbf{0.7--1.6} & \textbf{99.5} \\
\midrule
\multicolumn{22}{@{}l}{\emph{Qwen1.5-MoE-A2.7B, 60 experts}} \\
No balancing & & \multicolumn{3}{c}{57.5\hspace{1.15em}59.3} & \multicolumn{3}{c}{57.5\hspace{1.15em}59.3} & \multicolumn{3}{c}{57.5\hspace{1.15em}59.3} & \multicolumn{3}{|c}{65.1\hspace{1.15em}61.2} & \multicolumn{3}{c}{65.1\hspace{1.15em}61.2} & \multicolumn{3}{c}{65.1\hspace{1.15em}61.2} &  & 60.8 \\
EPLB (deployed) & & \multicolumn{3}{c}{68.7\hspace{1.15em}64.0} & \multicolumn{3}{c}{68.7\hspace{1.15em}64.0} & \multicolumn{3}{c}{68.7\hspace{1.15em}64.0} & \multicolumn{3}{|c}{70.5\hspace{1.15em}62.6} & \multicolumn{3}{c}{70.5\hspace{1.15em}62.6} & \multicolumn{3}{c}{70.5\hspace{1.15em}62.6} &  & 66.4 \\
LPLB (GPU LP)$^{a}$ & & \multicolumn{3}{c}{49.5\hspace{1.15em}$\times$} & \multicolumn{3}{c}{83.1\hspace{1.15em}$\times$} & \multicolumn{3}{c}{95.1\hspace{1.15em}$\times$} & \multicolumn{3}{|c}{49.5\hspace{1.15em}$\times$} & \multicolumn{3}{c}{83.1\hspace{1.15em}$\times$} & \multicolumn{3}{c}{95.1\hspace{1.15em}$\times$} & 102 & 75.9 \\
FineMoE (CPU LP)$^{a}$ & & \multicolumn{3}{c}{32.6\hspace{1.15em}20.8} & \multicolumn{3}{c}{70.7\hspace{1.15em}56.8} & \multicolumn{3}{c}{90.6\hspace{1.15em}84.0} & \multicolumn{3}{|c}{32.6\hspace{1.15em}20.8} & \multicolumn{3}{c}{70.7\hspace{1.15em}56.8} & \multicolumn{3}{c}{90.6\hspace{1.15em}84.0} & 207--380 & 59.3 \\
EPLB + relief (kernel), $I^{*}$ & & \multicolumn{3}{c}{86.6\hspace{1.15em}83.2} & \multicolumn{3}{c}{96.3\hspace{1.15em}94.5} & \multicolumn{3}{c}{98.9\hspace{1.15em}98.0} & \multicolumn{3}{|c}{87.1\hspace{1.15em}84.7} & \multicolumn{3}{c}{96.5\hspace{1.15em}94.9} & \multicolumn{3}{c}{99.0\hspace{1.15em}98.2} & 13--26 & 93.2 \\
EPLB + relief (FPGA), $I^{*}$ & & \multicolumn{3}{c}{97.0\hspace{1.15em}96.3} & \multicolumn{3}{c}{99.3\hspace{1.15em}99.1} & \multicolumn{3}{c}{99.8\hspace{1.15em}99.7} & \multicolumn{3}{|c}{97.1\hspace{1.15em}96.4} & \multicolumn{3}{c}{99.3\hspace{1.15em}99.1} & \multicolumn{3}{c}{99.8\hspace{1.15em}99.7} & 2.8--5.0 & 98.5 \\
EPLB + relief (silicon), $I^{*}$ & & \multicolumn{3}{c}{\textbf{99.0}\hspace{1.15em}\textbf{98.6}} & \multicolumn{3}{c}{\textbf{99.8}\hspace{1.15em}\textbf{99.6}} & \multicolumn{3}{c}{\textbf{99.9}\hspace{1.15em}\textbf{99.9}} & \multicolumn{3}{|c}{\textbf{99.1}\hspace{1.15em}\textbf{98.7}} & \multicolumn{3}{c}{\textbf{99.8}\hspace{1.15em}\textbf{99.6}} & \multicolumn{3}{c}{\textbf{99.9}\hspace{1.15em}\textbf{99.9}} & \textbf{0.7--2.0} & \textbf{99.5} \\
\bottomrule
\end{tabular}

\end{table*}

\subsection{Experimental Setup}
\label{sec:eval:method}

Routing is captured from three real MoE families on two
datasets each, GSM8K and WildChat. Qwen3-30B-A3B (128 experts, top-8, all 48
layers, 48 records per layer) provides the full group-size range, while OLMoE-1B-7B
(64 experts, top-8, 16 layers) and Qwen1.5-MoE-A2.7B (60 experts, top-4, 24
layers) repeat the comparison at the group sizes their expert counts admit. Placements are $D{=}2$ tori at
the power-of-two group sizes deployments use, $G \in \{8,16,32,64\}$
($C_2{\times}C_4$ up to $C_8{\times}C_8$), with two replica slots per rank, so a
replicated expert lives on exactly two ranks, its home and one peer. An expert count that does not divide $G$
is zero-padded to the next multiple, identically for every scheme. Every balancing scheme runs on
the same EPLB placement, computed once from each layer's mean load, and varies only the split. The
no-balancing baseline keeps the default placement with the same two redundant slots
per rank. The grouped scheme is
the deployment LPLB's own documentation recommends at scale. The eight-rank
template tiles across $G/8$ independent groups, each solving its own exact
program, at a solve latency measured on the same GPU. Every latency in
this section is measured on a GPU that admits 99\,KiB of dynamic shared
memory per thread block after opt-in.
The no-balancing and EPLB rows keep the production engines' oblivious
replica pick of Table~\ref{tab:tiers} as their split.
Fitting the ratios to the layer mean instead moves the EPLB row in either direction depending on the
configuration, so no fixed ratio, tuned or oblivious,
responds to the per-step residue. Every number attributed to the engine comes from the fixed-point reference
the RTL matches bit-exactly. Every aggregate is layer-major. Each of the 48 layers keeps its
own routing ensemble and its own EPLB placement, and cross-layer numbers are
means of per-layer results. Pooling records across layers instead fits the
placement to an average layer that appears in no trace and counts
inter-layer variance as per-step variance, which overstates exactly the
adaptation this paper advocates. The reference
for every comparison is the per-step optimum, the exact split re-solved every step
at zero cost. Each software solver's cost enters through Eq.~\ref{eq:ueff}, and its
launch limit enters as absence. Producing the placement itself is likewise charged to no scheme, a
concession of different size for the two mechanisms. EPLB's re-fit moves
expert weights, megabytes per migration, where a placement change costs
Relief one re-colouring, a 912-byte list written once
(\S\ref{sec:engine:uarch2}). Neither charge changes a figure in this
section. The engine's cycle counts come from RTL simulation on these traces, at
every group size and family, and are the same in every deployment. The
silicon deployment's area and timing come from the OpenROAD flow on the
ASAP7 7\,nm educational PDK~\cite{clarkASAP77nmFinFET2016}, quoted from the
clock-tree stage, after timing repair, where its analyzer reports a
2.89\,ns minimum period with no violated setup or hold path. The parasitics at that stage are global-route estimates. The FPGA deployment is placed and
routed to timing closure on its device. Every
figure charges the engine as fully blocking, with no overlap with the
layer's other work.

\subsection{End-to-End Results}
\label{sec:eval:headline}

Figure~\ref{fig:ueffpair} plots \ueff{} across \tmoe{} at every group
size. The silicon deployment attains the highest effective utilization of
any scheme at every point of every panel, and runs within 1\% of the
cost-free optimum once \tmoe{} exceeds a few hundred microseconds. The
closing paragraph of this subsection compares the three deployments against
one another. Where the exact GPU solver runs, its balance is optimal and its latency
still holds it below the no-balancing scheme at the shortest layers. Past
its launch limit it has no curve at all
(\S\ref{sec:model:premise}). The CPU implementation runs at every group size. Its latency keeps it below every per-step alternative, and
the same optimum computed off the GPU never becomes the best scheme in
the range.

Table~\ref{tab:verdict} reports every scheme and model family as \ueff{}
relative to the per-step optimum, at three operating points and the group
sizes each family admits.
They are the exact GPU solver's measured break-even at 100\,\textmu s, the
500\,\textmu s production decode scale that DeepEP's collectives
anchor~\cite{DeepseekaiDeepEP2026}, and a large-batch layer at
2000\,\textmu s where decision latency is negligible.
Figure~\ref{fig:ueffpair} carries every point between and beyond them. According to the table, \ueff{} orders the schemes differently
than balance does, since the GPU solver holds all of the attainable balance
yet delivers less utilization than leaving the split alone at the shortest
operating point. The engine's rows are the best deployable value in
every column of every family, where the 11.8 to 15.9 points of that share between engine and kernel at
100\,\textmu s are the shared-memory path expressed in utilization. 

The deployed static scheme's advantage over no balancing changes sign at the largest group. It measures $+16.7$, $+18.6$, $+9.4$, and $-3.5$ points of spatial
utilization across $G \in \{8,16,32,64\}$ on GSM8K, with WildChat replicating the
collapse at $+14.8$, $+13.0$, $+7.2$, $-1.7$, the reversal
\S\ref{sec:model:premise} isolates with a controlled probe. Relief's per-step split on that same replicated
placement reaches 0.28 to 0.29 depending on \tmoe{}, five points of balance over
the best deployable scheme at the scale where the exact GPU solver does not
run at all. The group-local solver is the closest competitor there, and it is the
configuration LPLB's documentation recommends for large deployments. Its eight-rank programs solve in a measured 33\,\textmu s and never meet
the launch limit. They cannot move tokens across group boundaries, so its
balance tops out at 91.6\% of the optimum's, and the two together leave it
at 68.7\% at the 100\,\textmu s operating point, below the no-balancing
row.

\begin{figure}[t]
 \centering
 \includegraphics[width=\linewidth]{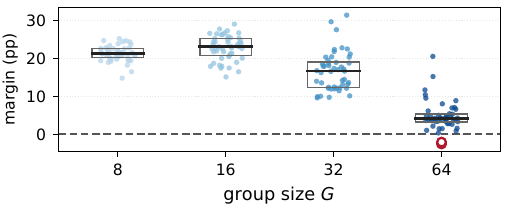}
 \caption{Spatial-utilization margin of Relief, at the production
 operating point's solved pass count, over the best other deployable
 scheme, one dot per layer of the 128-expert family at each group size. Boxes mark
 quartiles and medians, and the two open
 red points are the $G{=}64$ exceptions of \S\ref{sec:eval:headline}.
 Relief attains the higher utilization in all but those two of the 192
 combinations.}
 \label{fig:zeroreg}
\end{figure}

At its solved pass counts Relief's balance follows the layer it
serves. It stays within 0.17\% of the per-step optimum at $G{=}16$ and 0.18\% at
$G{=}64$ at the shortest operating point, tightening further
from the production scale on. At the production point's pass count it is
the best achievable scheme in 190 of 192 layer-group combinations
(Fig.~\ref{fig:zeroreg}). On two $G{=}64$ layers the default placement with no balancing at all wins
by 9 to 12\%, because the replicated placement fits that routing worse than
the default does. Relief still sits within 0.6\% of its own placement's optimum on
both layers, so what falls short there is the placement, which remains a
separate, composable mechanism (\S\ref{sec:engine}).
Figure~\ref{fig:zeroreg} plots every cell's margin over the best other
deployable scheme. The median across all combinations is 19.1\,pp and the maximum
31.4\,pp. The per-$G$ medians fall from the low twenties at $G{\in}\{8,16\}$
to 4.1\,pp at $G{=}64$ as intra-rank pooling thins, and the margin is
negative only in the two $G{=}64$ layers above.

\begin{figure}[t]
 \centering
 \includegraphics[width=\linewidth]{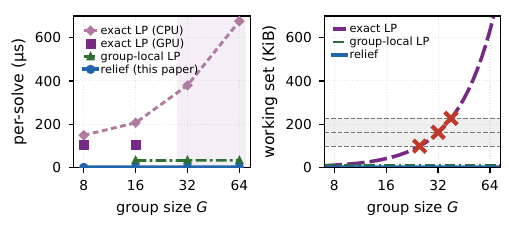}
 \caption{Per-solve time (left) and working set (right) against group size. The
 shaded region is where the GPU solver no longer launches on the measured
 GPU, and each red cross marks where its $156G^{2}$-byte working set
 crosses one of the shipping per-thread-block capacities the grey band
 spans, one cross each for the 5090, A100 and H100. The engine's own
 curves are flat against both axes.}
 \label{fig:hwscale}
\end{figure}

Per-solve time is nearly flat in group size (Fig.~\ref{fig:hwscale}),
0.75\,\textmu s at $G{=}8$ to 0.79 at $G{=}64$ for the silicon deployment.
The FPGA deployment sits a constant PCIe term above it. The margins therefore
persist as the deployment scales. The live working set grows at twelve bytes per rank, where the exact
solver's $156G^{2}$ bytes cross every shipping GPU's shared-memory capacity by
$G{=}39$. The expert count enters through the scan schedule alone
(\S\ref{sec:engine:uarch2}), so a 256- or 1024-expert model at $G{=}64$
lengthens the scan by a few chunks and changes no wire, port, or register
width.

The three deployments differ only in \tsolve{}, so every margin between
their rows is the interface geometry of \S\ref{sec:engine}.

\subsection{Choosing the Pass Count}
\label{sec:eval:dial}

The pass count of Eq.~\ref{eq:istar} follows the layer it serves.
$I^{*}$ spans 6 to 64 across the operating points and group sizes of
Table~\ref{tab:verdict}, because a longer layer absorbs more passes. Both hardware curves of
Fig.~\ref{fig:ueffpair} are drawn at the value it returns. Hardwiring $I{=}12$ instead costs up to 1.6 points of the optimum's share
in a cell of Table~\ref{tab:verdict}, and at most 0.7 of a point on a
family mean.

\begin{figure}[t]
 \centering
 \includegraphics[width=\linewidth]{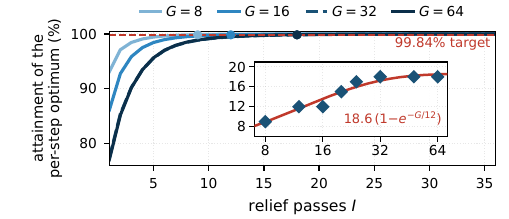}
 \caption{Attainment of the per-step optimum against pass count, at every
 integer $I$ from 1 to 36, one curve per group size, with a circle where each curve crosses the 99.84\% frontier. The inset
 gives that frontier's pass count (vertical) against group size
 (horizontal, diamonds) at eight sizes, the four
 deployment sizes plus four near-square tori ($G{=}12,20,24,48$) measured
 only for this scaling. The solid curve there is the saturating fit
 $18.6\,(1{-}e^{-G/12})$. Three of the eight sizes lie past the measured GPU's launch limit of $G{=}25$.}
 \label{fig:limitsmap}
\end{figure}

Over the full $(G, I)$ plane (Fig.~\ref{fig:limitsmap}) the 99.84\%
attainment frontier, measured at eight group sizes, rises from nine passes
to eighteen at $G{=}32$ and stays there. 

\subsection{Ablation}
\label{sec:eval:ablation}

Table~\ref{tab:ablation} compares the relaxation against its own variants
and against the rival relaxations, so each difference isolates the removed
part. Replacing the live loads with a pass-start snapshot makes the relaxation
embarrassingly parallel, a damped Jacobi pass, and removes the most
attainment. It loses 4.3 and 9.2 points, recovers Relief's balance only
at four times the passes, and diverges without the damping. Among the
live-read variants, removing the reverse move loses the most, since flow
then only ever leaves the heavier endpoint and a rank made hottest by
what it received can never be relieved. The compiled colouring order matches the adaptive walk at $G{=}16$ and
lands 0.12 points further at $G{=}64$. Every deployment walks the fixed
list that order produces. Water-filling and proportional
redistribution, two standard relaxations, land an order of magnitude
further from the optimum at an equal pass count. Outside the table, the
kernel takes 19--29$\times$ longer than the engine for the
same splits, the shared-memory path of \S\ref{sec:deploy:kernel} measured
at equal pass counts. The descriptor interface's $96\times$ transaction
reduction keeps the bus from dominating \tsolve{} at Relief's timescale
(\S\ref{sec:engine:iface}).

\begin{table}[t]
\centering\footnotesize
\setlength{\tabcolsep}{4pt}
\caption{Attainment of the per-step optimum under ablations of the
relaxation rule and of the arithmetic. Every variant runs at $I{=}12$ passes on the same placement, on the
128-expert family's GSM8K routing, layer-major. FP32 is the exact
LP's own precision and produces the reference optimum, so the LP appears
only in the fixed-point row.}
\label{tab:ablation}
\begin{tabular}{@{}lcc@{}}
\toprule
& \multicolumn{2}{c}{attainment of optimum (\%)} \\
\cmidrule(l){2-3}
Variant & $G{=}16$ & $G{=}64$ \\
\midrule
\multicolumn{3}{@{}l}{\emph{Relaxation rule}} \\
Relief, engine reference (Q11.20) & \textbf{99.84} & \textbf{99.51} \\
\quad without the reverse move & 98.80 & 98.43 \\
\quad snapshot loads, damped (Jacobi) & 95.55 & 90.35 \\
\midrule
\multicolumn{3}{@{}l}{\emph{Other relaxations, same pass count}} \\
Water-filling relaxation & 86.23 & 75.22 \\
Proportional relaxation & 93.09 & 81.04 \\
Hash split (deployed) & 72.54 & 70.80 \\
\midrule
\multicolumn{3}{@{}l}{\emph{Arithmetic}} \\
\quad relief in FP32 instead & 99.68 & 98.74 \\
\quad relief in Q11.8 & 99.83 & 99.49 \\
\quad relief in Q11.4 & 99.36 & 99.05 \\
\quad the exact LP in Q11.20 & 73.68 & 73.73 \\
\bottomrule
\end{tabular}

\end{table}

The engine computes in Q11.20 fixed point, a 32-bit signed word, the same
width as the FP32 solver it displaces, and the lower section of
Table~\ref{tab:ablation} reports every scheme's attainment under each
32-bit arithmetic. Relief's operands are bounded, same-scale integer
counts, which a fixed radix point covers exactly, so the fixed word matches
the float one. Moreover, the truncating transfer's deadband, by damping the
relaxation's small oscillations, even carries the fixed-point split
slightly past it at an equal pass count. The narrower rows show the
fractional field flat down to 8 bits. The exact LP does not survive the
same word, because an interior-point solve is precision-sensitive. Its
normal-equations factorization needs the range only a floating radix
spans, and in Q11.20 its attainment collapses to 73\%. The few tenths of a percent that Relief concedes to the optimum change no
ranking in
\S\ref{sec:eval:headline}.

\subsection{Hardware Overhead}
\label{sec:characterization}

The engine comes to 0.14\,mm$^2$ of standard cells in a 0.35\,mm$^2$
outline. Its state is 37{,}434 flip-flops. Its 1.29\,M cells are ten million transistors by the per-cell static-CMOS
count, 0.005--0.021\% of a current accelerator by transistor count and
0.017--0.022\% by area, or up to 0.06\% if the whole floorplan outline is
charged,
an order of magnitude below the 0.93\% Morpheus
reports for its on-die controller~\cite{darabiMorpheusExtendingLast2022}.
Absolute power is left unreported because vectorless activity estimation
diverges on this netlist's reconvergent arithmetic.

\begin{table}[t]
\centering\footnotesize
\setlength{\tabcolsep}{5pt}
\caption{The engine against the dies it would serve, and the FPGA deployment's
post-route occupation. Die figures are the vendors'. n/p: not published as
a single die. ASAP7 is a predictive 7\,nm PDK, less dense than the production nodes, so
the area shares are upper bounds and the transistor shares are
node-independent.}
\label{tab:overhead}
\begin{tabular}{@{}lrrrr@{}}
\toprule
& \multicolumn{2}{c}{absolute} & \multicolumn{2}{c}{engine share} \\
\cmidrule(lr){2-3}\cmidrule(l){4-5}
Block / accelerator die & mm$^2$ & trans. & by area & by trans. \\
\midrule
\textbf{Relief engine} (ASAP7) & \textbf{0.14} & \textbf{10\,M} & --- & --- \\
\midrule
A100 (7\,nm) & 826 & 54\,B & 0.02\% & 0.02\% \\
H100 (4\,nm) & 814 & 80\,B & 0.02\% & 0.01\% \\
RTX 5090 (4\,nm) & 750 & 92\,B & 0.02\% & 0.01\% \\
B200, per die (4\,nm) & n/p & 104\,B & --- & 0.01\% \\
AMD MI355X, package (3\,nm) & n/p & 185\,B & --- & 0.01\% \\
Tesla Dojo D1~\cite{talpesMicroarchitectureDOJOTeslas2023} (7\,nm) & 645 & 50\,B & 0.02\% & 0.02\% \\
Graphcore GC200~\cite{knowlesGraphcore2021} (7\,nm) & 823 & 59\,B & 0.02\% & 0.02\% \\
\midrule
& LUTs & FFs & DSPs & BRAM \\
\textbf{Relief engine} (XCVH1582) & 278{,}895 & 36{,}083 & 20 & 0 \\
\bottomrule
\end{tabular}

\end{table}

\section{Related Work}
\label{sec:related}

\textbf{Exact per-step splitting.} LPLB solves the min--max split as an
interior-point LP on the GPU~\cite{DeepseekaiLPLB2026}, and the same formulation balances
training micro-batches on CPU solvers~\cite{zhaoFinegrainedMoELoad2026}; MoETuner casts
placement itself as an offline ILP~\cite{goMoETunerOptimizedMixture2025}. Our measurements confirm that premise and quantify why software cannot
deliver it at decode. Relief
reaches the same balance without the program.

\textbf{Placement and replication.} EPLB re-places replicated experts from
smoothed statistics~\cite{DeepseekaiEPLB2026}, and CRAFT allocates replica counts under a
memory budget offline~\cite{zhaoCRAFTFineGrainedCostAware2026}. Both decide which expert copies exist, and Relief runs on whatever
placement they produce.

\textbf{Training-time shaping.} Auxiliary balance losses~\cite{lepikhinGShardScalingGiant2020, fedusSwitchTransformersScaling2022}, loss-free bias adjustment~\cite{wangAuxiliaryLossFreeLoadBalancing2024}, BASE's optimal
assignment~\cite{lewisBASELayersSimplifying2021}, and expert-choice
routing~\cite{zhouMixtureofExpertsExpertChoice2022} shape the router during training. They set the distribution the serving system receives, while Relief
balances, at every step, whatever distribution arrives.

\textbf{Device-level MoE serving.} Accelerator-side work restructures where
experts execute. Duplex and MoNDE move low-intensity or cold experts toward
memory~\cite{yunDuplexDeviceLarge2024, kimMoNDEMixtureNearData2024},
MoE-OPU compresses and prefetches on one
FPGA~\cite{luMoEOPUFPGAOverlay2025}, Pre-gated MoE and ProMoE prefetch
predicted experts~\cite{hwangPregatedMoEAlgorithmSystem2024, songProMoEFastMoEbased2025}, and Sem-MoE co-schedules
placement with token batching to shrink the all-to-all~\cite{liSEMANTICPARALLELISMREDEFINING2026}. A
parallel line schedules and overlaps the all-to-all collective
itself~\cite{hwangTutelAdaptiveMixtureofExperts2023, liAcceleratingDistributedMoE2023, liuJanusUnifiedDistributed2023}. They decide where experts execute or how fast tokens travel, and several
evaluate under the uniform routing our traces remove, so none of them
balances per-step dispatch and all of them compose with a split. Faster
collectives in fact raise the weight of the split's latency, since every
microsecond removed from \tmoe{} increases $\tsolve/\tmoe$ at any fixed
decision cost (Eq.~\ref{eq:ueff}).

\textbf{Hardware load balancers.} Balancing at sub-decision-cost granularity has
moved into hardware before, for reasons of the same kind. RPCValet steers
microsecond RPCs from an on-chip NI because an I/O-bus round trip makes decisions
stale~\cite{daglisRPCValetNIDrivenTailAware2019}; Turbo shows that per-request dispatch keeps its value even against free
periodic rebalancing~\cite{seyedroudbariTurboSmartNICenabledDynamic2023}; Intel ships a
fixed-function dynamic load balancer on Xeon dies~\cite{louDynamicLoadBalancer2025a}; and AWB-GCN
rebalances processing elements inside a sparse accelerator at
runtime~\cite{gengAWBGCNGraphConvolutional2020}. The engine balances GPUs in
an expert-parallel group, where RPCValet and the Intel block balance cores
behind a NIC and AWB-GCN balances lanes inside one accelerator.
Re-placement, offline planning, and single-device execution remain the
domain of the systems above.

\section{Conclusion}
\label{sec:conclusion}

Expert-parallel MoE serving wastes most of its expert-stage GPU time on
stragglers, and recovering it takes one balancing decision per layer per
decode step, a rate no software solver sustains. Relief comes within a fraction of a point of the per-step optimum in a few
passes of integer arithmetic, in a kernel that outruns the solver and in a fixed-function form that adds 0.02\% of an accelerator's transistors. Across three model families it matches or exceeds every deployable
scheme's effective utilization, recovering up to 37.3 points of that share.

\bibliographystyle{IEEEtranS}
\bibliography{refs}

\appendix
\section*{AI Use}

The authors used Claude (Anthropic) to improve the clarity of the paper. All technical content and ideas are the authors' work.

\end{document}